\documentclass[conference,letterpaper]{IEEEtran}

\usepackage[utf8]{inputenc} 
\usepackage[T1]{fontenc}
\usepackage{url}
\usepackage{ifthen} 
\newboolean{short_version}
\setboolean{short_version}{false} 
\usepackage[cmex10]{amsmath} 

\usepackage[url,hyperrefblack,notheorems,IEEEtran]{research17} 
\usepackage{balance} 
\usepackage{amssymb,amsfonts,amsbsy}
\usepackage{enumitem}
\usepackage{mathtools}
\usepackage[lined,boxed,commentsnumbered,linesnumbered,ruled]{algorithm2e}
\usepackage{comment}
\setlist[description]{leftmargin=\parindent,labelindent=\parindent}

\newtheorem{theorem}{\mytheoremname}
\newtheorem{lemma}{\mylemmaname}
\newtheorem{corollary}{\mycorollaryname}

\newtheorem{proposition}{\mypropositionname}

\newtheorem{definition}{\mydefinitionname}

\newtheorem{example}{\myexamplename}
\newcommand{\Hwt}[1]{\wH\left(#1\right)} 
\newcommand{\vol}[1]{\operatorname{vol}\left(#1\right)} 
\newcommand{\dual}[1]{{#1}^\perp} 
\newcommand{\ConstrA}[1]{\Lambda_\textnormal{A}(#1)} 
\newcommand{\ConstrAfour}[1]{\Lambda_{\textnormal{A}_4}(#1)} 
\newcommand{\ConstrC}[2]{\Lambda_{\textnormal{C}}(#1,#2)} 
\newcommand{\PConstrC}[2]{\Gamma_{\textnormal{C}}(#1,#2)} 
\newcommand{\we}[1]{W_{#1}} 
\newcommand{\jwe}[2]{\textnormal{jwe}_{#1,#2}} 
\newcommand{\swe}[1]{\textnormal{swe}_{#1}} 

\newcommand*{\Scale}[2][4]{\scalebox{#1}{\ensuremath{#2}}} 
\makeatletter
\renewcommand*\env@matrix[1][*\c@MaxMatrixCols c]{%
  \hskip -\arraycolsep
  \let\@ifnextchar\new@ifnextchar
  \array{#1}}
\makeatother
\usepackage{tikz}
\usetikzlibrary{calc,shapes, patterns,decorations.text, decorations.pathreplacing}
\usepackage{ctable} 
\usepackage{afterpage} 
\usepackage{lscape} 
\usepackage{pdflscape} 
\usepackage{rotating} 
\usepackage{pbox} 

\usepackage{todonotes}

\begin{document}
\title{On the Secrecy Gain of Formally Unimodular Construction $\textnormal{A}_4$ Lattices 
} 


\author{%
 \IEEEauthorblockN{Maiara F.~Bollauf, Hsuan-Yin Lin, and {\O}yvind Ytrehus}
 \IEEEauthorblockA{Simula UiB, N--5006 Bergen, Norway\\             
             Emails: \{maiara, lin, oyvindy\}@simula.no}
}

\maketitle

\begin{abstract}
  Lattice coding for the Gaussian wiretap channel is considered, where the goal is to ensure reliable communication between two authorized parties while preventing an eavesdropper from learning the transmitted messages. Recently, a measure called \emph{secrecy gain} was proposed as a design criterion to quantify the secrecy-goodness of the applied lattice code. In this paper, the theta series of the so-called \emph{formally unimodular lattices} obtained by Construction~$\textnormal{A}_4$ from codes over $\Integers_4$ is derived, and we provide a universal approach to determine their secrecy gains. Initial results indicate that Construction $\textnormal{A}_4$ lattices can achieve a higher secrecy gain than the best-known formally unimodular lattices from the literature. Furthermore, a new code construction of formally self-dual $\Integers_4$-linear codes is presented.
\end{abstract}


\section{Introduction}
\label{sec:introduction}

The study of \emph{physical layer security (PLS)} has recently received significant attention in the 5G and beyond 5G (B5G) wireless communications~\cite{WuKhisti-etal18_1,Bloch-etal21_1}. In contrast to cryptographic algorithms, approaches in PLS only utilize the resources at the physical layer of the transmitting parties and provide \emph{information-theoretically unbreakable security}. It stemmed from Aaron D.~Wyner's landmark paper~\cite{Wyner75_1} in 1975, which showed that based on the \emph{communication channel} characteristics, one can achieve communication that is reliable and at the same time secure against an adversarial eavesdropper.

In the famous \emph{wiretap channel (WTC)} introduced in~\cite{Wyner75_1}
, a single transmitter (Alice) tries to communicate to a receiver (Bob) while keeping the transmitted messages secure from an unauthorized eavesdropper (Eve). 
The secure and confidential achievable rate between Alice and Bob for WTC is defined as the \emph{secrecy rate}. There is a recent focus on designing practical wiretap codes that achieve a high secrecy rate based on lattices over Gaussian WTCs~\cite{BelfioreOggier10_1,OggierSoleBelfiore16_1,LingLuzziBelfioreStehle14_1}. Among these works, one of the essential design criteria for good wiretap lattice codes is the \emph{secrecy gain}~\cite{BelfioreOggier10_1,OggierSoleBelfiore16_1}, which is defined as the maximum attainable \emph{secrecy function} (the coding gain of a specifically designed lattice $\Lambda_\textnormal{e}$ for Eve compared to a regular integer lattice, evaluated in terms of the \emph{theta series} of lattices. See Section~\ref{sec:secrecy-function_lattice} for an explicit definition.)

Another design criterion for wiretap lattice codes, called the \emph{flatness factor}, was proposed by Ling \emph{et al.}~\cite{LingLuzziBelfioreStehle14_1}. The flatness factor quantifies how much confidential information can leak to Eve in terms of mutual information, while the secrecy gain characterizes Eve's success probability of correctly guessing the transmitted messages. The secrecy gain and the flatness factor both require small theta series of the designed Eve's lattice $\Lambda_\textnormal{e}$ at a particular point to guarantee secrecy-goodness~\cite{LingLuzziBelfioreStehle14_1}.

In this work, the quality criterion of secrecy gain for lattice coding is considered. Secrecy gains of the so-called \emph{unimodular} lattices have been studied for well  over a decade~\cite{BelfioreOggier10_1}. In this pioneering work, Belfiore and Sol{\'{e}} discovered that there exists a symmetry point in their secrecy functions. Further, they conjectured that for unimodular lattices, the secrecy gain is achieved at the symmetry point of its secrecy function. The conjecture has been further investigated and verified for unimodular (or \emph{isodual}) lattices in dimensions less than $80$~\cite{OggierSoleBelfiore16_1,Ernvall-Hytonen12_1,LinOggier13_1,Pinchak13_1}. The study of secrecy gain was recently also extended to the \emph{$\ell$-modular lattices}~\cite{OggierSoleBelfiore16_1,LinOggierSole15_1,OggierBelfiore18_1}, where it is believed that the higher the parameter $\ell$ is, the better secrecy gain we can achieve. Most recently, a new family of lattices, called \emph{formally unimodular lattices}, or lattices with the same theta series as their dual, was introduced~\cite{BollaufLinYtrehus22_1app}.\footnote{
  A \emph{formally-self dual} code has the same weight enumerator as its dual.} It was shown that formally unimodular lattices have the same symmetry point as unimodular and isodual lattices, and the Construction A lattices obtained from the formally self-dual codes can achieve a higher secrecy gain than the unimodular lattices. \ifthenelse{\boolean{short_version}}{}{{Moreover, for formally unimodular lattices obtained by Construction A from even formally self-dual codes, a sufficient condition to verify Belfiore and Sol{\'{e}}'s conjecture on the secrecy gain is also provided. (An \emph{even} code has all of its codewords with even weights. Otherwise, the code is \emph{odd}.)}}

This paper especially focuses on the analysis of the secrecy gain for formally unimodular lattices obtained by Construction $\textnormal{A}_4$ from codes over the ring $\Integers_4\eqdef\{0,1,2,3\}$ (also called the \emph{quaternary codes})~\cite{ConwaySloane93_1,BonnecazeSoleCalderbank95_1,Wan97_1}. Our contributions are three-fold:
\begin{enumerate}[nosep,label=\roman*)]
\item A code $\code{C}$ over $\Integers_4$ is \emph{formally self-dual} if it has 
the same \emph{symmetrized weight enumerator (swe)} as its dual. We show that if $\code{C}$ is formally self-dual, then its corresponding Construction $\textnormal{A}_4$ lattice is formally unimodular.
\item We provide a novel and universal approach to determine the secrecy gain for Construction $\textnormal{A}_4$ lattices obtained from formally self-dual codes over $\Integers_4$.
\item To study secrecy gains of Construction $\textnormal{A}_4$ lattices from formally self-dual $\Integers_4$-linear codes, we present a new code construction of formally self-dual $\Integers_4$-linear codes with respect to swe. There is not much known about this in the literature~\cite{GulliverHarada01_1,BetsumiyaHarada03_1,YooLeeKim17_1}.
\end{enumerate}

To the best of our knowledge, most of the efforts to solve Belfiore and Sol{\'{e}}'s conjecture on the secrecy gain of formally unimodular lattices have only been based on the lattices obtained by Construction A from binary codes. The investigation of Construction $\textnormal{A}_4$ lattices obtained from formally self-dual codes over $\Integers_4$ has not been addressed in the previous literature. \ifthenelse{\boolean{short_version}}{Due to page limitations, some proofs are omitted and can be found in the extended version~\cite{BollaufLinYtrehus22_1sub}.}{}

\section{Definitions and Preliminaries}
\label{sec:definitions-preliminaries}

\subsection{Notation}
\label{sec:notation}

We denote by $\Integers$, $\Rationals$, and $\Reals$ the set of integers, rationals, and reals, respectively. $[m:n]\eqdef\{m,m+1,\ldots,n\}$ for $m,n\in \Integers$, $m \leq n$. Vectors are boldfaced, e.g., $\vect{x}$. Matrices and sets are represented by capital sans serif letters and calligraphic uppercase letters, respectively, e.g., $\mat{X}$ and $\set{X}$. An identity matrix of dimensions $m \times m$ is denoted as $\mat{I}_{m}$, and $\mat{O}_{m\times n}$ represents an all-zero matrix of size $m\times n$. Denote by $d_{\textnormal{Lee}}(\vect{x},\vect{y})$ the \emph{Lee distance} between two vectors $\vect{x},\vect{y}$ over binary field $\Field_2$ or $\Integers_4$. $\inner{\vect{x}}{\vect{y}}$ denotes the inner product and the $\vect{x}\circ\vect{y}$ represents the element-wise (Hadamard/Schur) product between two vectors over $\Field_2$ or $\Integers_4$, respectively. We use the code parameters $[n,M]$ or $[n,M,d_{\textnormal{Lee}}]$ to denote a linear code $\code{C}$ of length $n$, $M$ codewords, and minimum \emph{Lee distance} $d_{\textnormal{Lee}}\eqdef\min_{\vect{x},\vect{y}\in \code{C}} d_{\textnormal{Lee}}(\vect{x},\vect{y})$. $\Hwt{\vect{x}}$ denotes the Hamming weight of a vector $\vect{x}$. $\phi_q\colon\Integers_q \rightarrow \Integers$ is defined as the natural embedding, i.e., $\phi_q(x)$ is the remainder of the division of $x$ by $q$. In this work, $q$ can be $2$ or $4$.

\subsection{Basics on Codes and Lattices}
\label{sec:basics_codes-lattices}

We next recall some definitions of codes over $\Field_2$, codes over $\Integers_4$, and lattices.

Let $\code{A}$ be an $[n,M]$ binary code. Its \emph{weight enumerator} is 
\begin{IEEEeqnarray*}{c}
  W_{\code{A}}(x,y)=\sum_{\vect{c}\in\code{A}} x^{n-\Hwt{\vect{c}}}y^{\Hwt{\vect{c}}}.
\end{IEEEeqnarray*}

Let $\code{C}_1,\code{C}_2$ be two binary linear codes. For $\vect{c}_1=(c_{1,1},\ldots,c_{1,n}) \in \code{C}_1, \vect{c}_2=(c_{2,1},\ldots,c_{2,n}) \in \code{C}_2$, we define 
\begin{IEEEeqnarray}{rCl}
  \IEEEyesnumber\label{eq:def_dij}
  \IEEEyessubnumber*
  d_{0,0}(\vect{c}_1,\vect{c}_2)& \eqdef & \card{\{j\in [1:n]\colon (c_{1,j},c_{2,j}) = (0,0)\}}, 
  \\ [1mm]
  d_{0,1}(\vect{c}_1,\vect{c}_2) & \eqdef & \card{\{j\in [1:n]\colon (c_{1,j}, c_{2,j})=(0,1)\}},
  \\[1mm]
  d_{1,0}(\vect{c}_1,\vect{c}_2) & \eqdef & \card{\{j\in [1:n]\colon (c_{1,j},c_{2,j}) = (1,0)\}},
  \\[1mm]
  d_{1,1}(\vect{c}_1,\vect{c}_2) & \eqdef &\card{\{j\in [1:n]\colon (c_{1,j}, c_{2,j})=(1,1)\}}.
\end{IEEEeqnarray}
Observe that $d_{0,0}(\vect{c}_1,\vect{c}_2) + d_{0,1}(\vect{c}_1,\vect{c}_2) + d_{1,0}(\vect{c}_1,\vect{c}_2)+ d_{1,1}(\vect{c}_1,\vect{c}_2) = n$. 
The \emph{joint weight enumerator} of $\code{C}_1$ and $\code{C}_2$ is given by
\begin{IEEEeqnarray}{c}
\textnormal{jwe}_{\code{C}_1,\code{C}_2}(a,b,c,d)\eqdef\sum_{\vect{c}_1 \in \code{C}_1} \sum_{\vect{c}_2 \in \code{C}_2} a^{d_{0,0}} b^{d_{0,1}} c^{d_{1,0}} d^{d_{1,1}},
\end{IEEEeqnarray}
where we use the shorthand $d_{i,j}$ for $d_{i,j}(\vect{c}_1,\vect{c}_2)$ defined in~\eqref{eq:def_dij}. Detailed properties and MacWilliams identities of the joint weight enumerator can be found in~\cite[Ch.~5, pp.~147--149]{MacWilliamsSloane77_1}.

A \emph{${\Integers_4}$-linear code} of length $n$ is an additive subgroup of $\mathbb{Z}_4^n$. 
If $\code{C}$ is a $\mathbb{Z}_4$-linear code of length $n$, then $\code{C}^\perp\eqdef\{{\bm x} \in \mathbb{Z}_4^n\colon \inner{\vect{x}}{\vect{y}} = 0,\,\textnormal{for all }\vect{y} \in \code{C}\}$ is the \emph{dual code} of $\code{C}$. 
    
From~\cite[Prop.~1.1]{Wan97_1}, it is well-known that any $\mathbb{Z}_4$-linear code is \emph{permutation equivalent} to a code $\code{C}$ with a generator matrix $\mat{G}$ in \emph{standard form}
\begin{IEEEeqnarray}{C}
  \mat{G}=\begin{pmatrix}
    \mat{I}_{k_1} & \mat{A} & \mat{B} \\
    \mat{O}_{k_1 \times k_2} & 2\mat{I}_{k_2} & 2\mat{C}
    \label{eq:Generator_Matrix}
  \end{pmatrix},
\end{IEEEeqnarray}
where $\mat{A}$ and $\mat{C}$ are binary matrices, and $\mat{B}$ is defined over $\Integers_4$. Such code $\code{C}$ is said to be a code of \emph{type $4^{k_1}2^{k_2}$}. 

The \emph{symmetrized weight enumerator (swe)} of a $\mathbb{Z}_4$-linear code $\code{C}$ is defined as
\begin{IEEEeqnarray*}{c}
  \textnormal{swe}_\code{C}(a,b,c)=\sum_{\vect{c}\in\code{C}} a^{n_0(\vect{c})}b^{n_1(\vect{c})+n_3(\vect{c})}c^{n_2(\vect{c})},
\end{IEEEeqnarray*}
where $n_i(\vect{c})\eqdef\card{\{j\in [1:n]\colon c_j=i\}}$, $i\in\Integers_4$. The corresponding MacWilliams identity for $\Integers_4$-linear codes is given by~\cite[Th.~2.3]{Wan97_1}
\begin{IEEEeqnarray}{rCl}
  \IEEEeqnarraymulticol{3}{l}{%
    \textnormal{swe}_{\code{C}}(a,b,c)}\nonumber\\*\quad%
  & = &\frac{1}{\card{\dual{\code{C}}}}\textnormal{swe}_{\dual{\code{C}}}(a+2b+c,a-c,a-2b+c). 
  \label{eq:swe-MacWilliams-identity_Z4}
\end{IEEEeqnarray}

Following the notion of swe, we have the following families of codes over $\Integers_4$.
\begin{definition}[Self-dual, isodual, formally self-dual codes]
  \begin{itemize}
  \item If $\code{C}=\dual{\code{C}}$, $\code{C}$ is a \emph{self-dual} code.
  \item If there is a permutation $\pi$ of coordinates such that $\code{C}=\pi(\dual{\code{C}})$, $\code{C}$ is called \emph{isodual}.
  \item If $\code{C}$ and $\dual{\code{C}}$ have the same symmetrized weight enumerator, i.e., $\textnormal{swe}_\code{C}(a,b,c)=\textnormal{swe}_{\dual{\code{C}}}(a,b,c)$, $\code{C}$ is a \emph{formally self-dual} code.
  \end{itemize}
\end{definition}
From~\eqref{eq:swe-MacWilliams-identity_Z4}, we can conclude that a code in any of these classes has its swe satisfying
\begin{IEEEeqnarray}{rCl}
  \IEEEeqnarraymulticol{3}{l}{%
    \textnormal{swe}_{\code{C}}(a,b,c)}\nonumber\\*\quad%
  & = &\frac{1}{\card{{\code{C}}}}\textnormal{swe}_{{\code{C}}}(a+2b+c,a-c,a-2b+c).
  \label{eq:swe-MacWilliams-identity_FSD-codes_Z4}
\end{IEEEeqnarray}

A (full rank) \emph{lattice} $\Lambda\subset\Reals^n$ is a discrete additive subgroup of $\mathbb{R}^{n}$, and it can be seen as
$\Lambda=\{\vect{\lambda}=\vect{u}\mat{L}_{n\times n}\colon\vect{u}\in\Integers^n\}$,
where the $n$ rows of $\mat{L}$ form a lattice basis in $\mathbb{R}^n$. 
The \emph{volume} of $\Lambda$ is $\vol{\Lambda} = \ecard{\det(\mat{L})}$. If a lattice $\Lambda$ has generator matrix $\mat{L}$, then the lattice $\Lambda^\star\subset\mathbb{R}^n$ generated by  $\trans{\bigl(\inv{\mat{L}}\bigr)}$ is called the \emph{dual lattice} of $\Lambda$. For lattices, the analogue of the weight enumerator of a code is the \emph{theta series}, defined as follows.
\begin{definition}[Theta series]
  \label{def:theta-series}
  Let $\Lambda$ be a lattice, its \emph{theta series} is given by
  \begin{IEEEeqnarray*}{c}
    \Theta_\Lambda(z) = \sum_{{\bm \lambda} \in \Lambda} q^{\norm{\vect{\lambda}}^2},
  \end{IEEEeqnarray*}
  where $q\eqdef e^{i\pi z}$ and $\Im{z} > 0$. 
\end{definition}

Analogously, the spirit of the MacWilliams identity can be captured by the \emph{Jacobi's formula}~\cite[eq.~(19), Ch.~4]{ConwaySloane99_1}
\begin{IEEEeqnarray}{c}
  \Theta_{\Lambda}(z)=\vol{\Lambda^\star}\Bigl(\frac{i}{z}\Bigr)^{\frac{n}{2}}\Theta_{\Lambda^\star}\Bigl(-\frac{1}{z}\Bigr).
  \label{eq:Jacobi-formula}
\end{IEEEeqnarray}

In some particular cases, the theta series of a lattice can be expressed in terms of the \emph{Jacobi theta functions} defined as follows.
\begin{IEEEeqnarray*}{rCl}
  \vartheta_2(z)& \eqdef &\sum_{m\in\Integers} q^{\bigl(m+\frac{1}{2}\bigr)^2}=\Theta_{\mathbb{Z} + \frac{1}{2}}(z),
  \nonumber\\
  \vartheta_3(z)& \eqdef &\sum_{m \in \mathbb{Z}} q^{m^2}=\Theta_{\mathbb{Z}}(z), ~ ~ \vartheta_4(z) \eqdef \sum_{m\in \mathbb{Z}} (-q)^{m^2}.
\end{IEEEeqnarray*}

Lattices can be classified according to their properties. It is said to be \emph{integral} if the inner product of any two lattice vectors is an integer. An integral lattice such that $\Lambda = \Lambda^\star$ is called a \emph{unimodular} lattice. A lattice $\Lambda$ is called \emph{isodual} if it can be obtained from its dual $\Lambda^\star$ by (possibly) a rotation or reflection. In~\cite{BollaufLinYtrehus22_1app}, a new and broader family was presented, namely the \emph{formally unimodular lattices}, that consists of lattices having the same theta series as their dual, i.e., $\Theta_{\Lambda}(z)=\Theta_{\Lambda^\star}(z)$.

Lattices can be constructed from binary linear codes through the so-called Constructions A and C~\cite{ConwaySloane99_1}.
\begin{definition}[Construction A]
  Let $\code{C}$ be a binary $[n,M]$ code, then $\ConstrA{\code{C}}\eqdef\frac{1}{\sqrt{2}}(\phi_2(\code{C}) + 2\Integers^n)$ is a lattice.
\end{definition}

\begin{definition}[$2$-level Construction C]
  \label{def:def_ConstrC}
  Let $\code{C}_1,\code{C}_2$ be two linear codes over $\Field_2$ and $\code{C}_1 \subseteq \code{C}_2$. If the chain $\code{C}_1 \subseteq \code{C}_2$ is closed under the element-wise product,\footnote{The chain $\code{C}_1 \subseteq \code{C}_2$ is called \emph{closed under the element-wise product} if for all $\vect{c}_1,\vect{c}'_1\in \code{C}_1$, we have $\vect{c}_1\circ \vect{c}'_1\in\code{C}_{2}$.} then the packing given by
  \begin{IEEEeqnarray}{c}
    \ConstrC{\code{C}_1}{\code{C}_2}\eqdef\phi_2(\code{C}_1) + 2\phi_2(\mathcal{C}_2) + 4\Integers^n
    \label{eq:ConstrC}
  \end{IEEEeqnarray}
  generates a lattice.
\end{definition}


For general choices of $\code{C}_1$ and  $\code{C}_2$, \eqref{eq:ConstrC} is a nonlattice packing, and we will denote by $\PConstrC{\code{C}_1}{\code{C}_2}$. If $\code{C}_1$ is the zero code, and $\code{C}_2$ is linear, then $\ConstrC{\code{C}_1}{\code{C}_2}=2\ConstrA{\code{C}_2}$. If $\code{C}_2$ is the universe code $\mathbb{F}_2^n$ and $\code{C}_1$ is linear, then $\ConstrC{\code{C}_1}{\code{C}_2}=\ConstrA{\code{C}_1}$. 
    
A packing $\Gamma\subset\mathbb{R}^n$ is \emph{geometrically uniform} if for any two elements ${\bm c},{\bm c'} \in \Gamma$ there exists an isometry $T$ such that ${\bm c'}=T({\bm c})$ and $T(\Gamma)=\Gamma$. It was demonstrated that $\PConstrC{\code{C}_1}{\code{C}_2}$ is geometrically uniform~\cite{BollaufZamir16_1}, for linear codes $\code{C}_1$ and $\code{C}_2$.

There is an analogue of Construction A for codes over $\mathbb{Z}_4$, which is called \emph{Construction $\textnormal{A}_4$}.
\begin{definition}[{Construction $\textnormal{A}_4$~\cite[Ch.~12.5.3]{HuffmanPless03_1}}]
  \label{def:def_ConstrAfour}
  If $\code{C}$ is a $\mathbb{Z}_4$-linear code, then $\ConstrAfour{\code{C}}=\frac{1}{2}(\phi_4(\code{C})+4\mathbb{Z}^n)$ is a lattice.
\end{definition}


It is known that $\ConstrAfour{\code{C}}$ is a unimodular lattice if and only if the $\Integers_4$-linear code $\code{C}$ is a self-dual code~\cite[Prop.~12.2]{Wan97_1}. For notational convenience, sometimes the mapping $\phi_q$ is omitted. 

\ifthenelse{\boolean{short_version}}{}{{
The following example illustrates Construction $\textnormal{A}_4$ of the octacode.
\begin{example}
\label{ex:E8_octacode}
The self-dual $\Integers_4$-linear code, known as the octacode $\code{O}_8$, is generated by $\mat{G} = (\mat{I}_{4} ~ \mat{B})$, where 
\begin{IEEEeqnarray*}{c}
\mat{B} = \begin{pmatrix}
3 & 1 & 2 & 1 \\
1 & 2 & 3 & 1 \\
3 & 3 & 3 & 2 \\
2 & 3 & 1 & 1
\end{pmatrix}.  
\end{IEEEeqnarray*}
It is of type $4^4$ and its swe~\cite[Ex.~12.5.13]{HuffmanPless03_1} is given by
\begin{IEEEeqnarray*}{rCl}
  \IEEEeqnarraymulticol{3}{l}{%
    \textnormal{swe}_{\code{O}_8}(a,b,c)}\nonumber\\*\,%
  & = &a^8+ 16a^8+ b^8+ 14a^4c^4+ 112a^3 b^4y + 112a b^4 c^3.
  \label{eq:swe_octacode}\IEEEeqnarraynumspace
\end{IEEEeqnarray*}
A unimodular lattice can be constructed by performing $\widebar{\lattice{E}}_8=\Lambda_{\textnormal{A}_4}(\code{O}_8) = \frac{1}{2}\left(\code{O}_8+4\mathbb{Z}^8\right)$. It is equivalent to the well-known Gosset lattice $\lattice{E}_8$~\cite[Ex.~12.5.13]{HuffmanPless03_1}. Note that the theta series of the $\lattice{E}_8$ lattice in terms of the Jacobi theta functions is
\begin{IEEEeqnarray*}{c}
\Theta_{\lattice{E}_8} = \tfrac{1}{2} \left (\vartheta_2(z)^8 + \vartheta_3(z)^8 + \vartheta_4(z)^8 \right).
\end{IEEEeqnarray*}

The $\lattice{E}_8$ lattice can also be constructed via the binary Construction A, using the $[8,4,4]$ extended Hamming code.\hfill\exampleend
\end{example}
}}

\section{Lattices from $2$-Level Construction C and Construction $\textnormal{A}_4$}
\label{sec:lattices_2-level-ConstrC-ConstructionAfour}

Some $\mathbb{Z}_4$-linear codes can be obtained from binary linear codes by using the $2$-level Construction C as in Definition~\ref{def:def_ConstrC}.

\begin{proposition}[{\cite[Lemma 2.1]{BonnecazeSoleCalderbank95_1}}]
  \label{prop:Z4-linear_ConsC}
  Consider two binary linear codes $\code{C}_1,\code{C}_2$, and let $\code{C}=\code{C}_1+2\code{C}_2\eqdef\{\vect{c}_1+2\vect{c}_2\colon\vect{c}_1\in\code{C}_1,\vect{c}_2\in\code{C}_2\}$. Then, the code $\code{C}$ over $\Integers_4$ is linear if and only if $\code{C}_1 \subseteq \code{C}_2$ is closed under the element-wise product.
\end{proposition}

On one hand, the condition that the chain $\code{C}_1 \subseteq \code{C}_2$ is closed under Schur product guarantees that  $\ConstrC{\code{C}_1}{\code{C}_2}$ is a lattice, on the other hand, $\code{C}$ being  $\Integers_4$-linear assures that $\ConstrAfour{\code{C}}$ is a lattice. Therefore, Proposition~\ref{prop:Z4-linear_ConsC} standardize the $2$-level Construction C and Construction $\textnormal{A}_4$, together with their respective conditions to be a lattice.

\ifthenelse{\boolean{short_version}}{}{{
Denote by $\code{C}_1$ an $[n,2^{k_1}]$ code and $\code{C}_2$ an $[n,2^{k_2}]$ code. Once a $\Integers_4$-linear code $\code{C}$ can be expressed as $\code{C}_1+2\code{C}_2$, and the codes $\code{C}_1$ and $\code{C}_2$ are generated, respectively, by
\begin{IEEEeqnarray*}{c}
  \mat{G}_1 =
  \begin{pmatrix}
    \mat{I}_{k_1} & \mat{X} & \mat{Y}
  \end{pmatrix},\quad
  \mat{G}_2 =
  \begin{pmatrix}
    \mat{I}_{k_1} &\mat{X} & \mat{Y} \\
    \mat{O}_{k_1 \times (k_2-k_1)} & \mat{I}_{k_2-k_1} & \mat{Z}
  \end{pmatrix}.
\end{IEEEeqnarray*}
Then, the generator matrix $\mat{G}$ of $\code{C}$ as in~\eqref{eq:Generator_Matrix} becomes~\cite[Thm.~3]{ConwaySloane93_1}
\begin{IEEEeqnarray}{c}
  \mat{G}=\begin{pmatrix}
    \mat{I}_{k_1} & \mat{X} & \mat{Y} \\
    \mat{O}_{k_1 \times (k_2-k_1)} & 2\mat{I}_{k_2-k_1} & 2\mat{Z}
    \label{eq:Generator_Matrix_C1C2}
  \end{pmatrix}.
\end{IEEEeqnarray}
}}
Up to now, we revised some results on the construction of $\Integers_4$-linear codes from two binary linear codes $\code{C}_1$ and $\code{C}_2$, and we notice that the lattice derived from them via Construction $\textnormal{A}_4$ (or analogously $2$-level Construction C) can have some properties, such as being unimodular or isodual.

\subsection{Weight Enumerators and Theta Series}
\label{sec:weight-enumerators_theta-series}

We now define a few extra notions of weight enumerators and derive an expression for the theta series of the lattice generated via Construction $\textnormal{A}_4$, given the symmetric weight enumerator of the $\Integers_4$-linear code $\code{C}$. An analogous relation will be discussed for the joint weight enumerator as well.


From the fact that a $2$-level Construction C is geometrically uniform for binary linear codes $\code{C}_1$ and $\code{C}_2$ together with the expression of the theta series of periodic packings given in~\cite{OdlyzkoSloane80_1}, we can state the following result. 
    
\begin{theorem}
  \label{thm:theta-series_2-level-constructionC}
  Consider a $2$-level Construction C lattice given by $\PConstrC{\code{C}_1}{\code{C}_2} = \tfrac{1}{2}(\code{C}_1 + 2\code{C}_2 + 4\mathbb{Z}^n),$ where $\code{C}_1, \code{C}_2$ are binary linear codes. The theta series of $\PConstrC{\code{C}_1}{\code{C}_2}$ is
  \begin{IEEEeqnarray*}{rCl}
    \IEEEeqnarraymulticol{3}{l}{%
      \Theta_{\PConstrC{\code{C}_1}{\code{C}_2}}(z)}\nonumber\\*\quad%
    & = &\sum_{\vect{c}_1 \in \code{C}_1} \sum_{\vect{c}_2 \in \code{C}_2} \vartheta_3^{d_{0,0}}(4z) \left(\frac{\vartheta_{2}(z)}{2}\right)^{{d_{1,0}} + {d_{1,1}}}  \vartheta_2^{{d_{0,1}}}(4z).
  \end{IEEEeqnarray*}  
\end{theorem}

\ifthenelse{\boolean{short_version}}{}{{
\begin{IEEEproof}
  Recall that the theta function of periodic packings, not necessarily lattices, can be obtained as follows.
  \begin{proposition}[\cite{OdlyzkoSloane80_1}]
    \label{prop:thm_thetanonl} 
    Given a periodic constellation $\Gamma = \bigcup_{k=1}^{M} (\Lambda + \vect{u}_k)$, where $\Lambda \subset \mathbb{R}^n$  is a lattice and ${\bm u}_1, \dots, {\bm u}_M \in \mathbb{R}^n$ are the  $M$ coset representatives. Then
    \begin{IEEEeqnarray}{c}
      \Theta_\Gamma(z) = \Theta_\Lambda(z) + \dfrac{2}{M} \displaystyle\sum_{k< \ell}\displaystyle\sum_{{\bm \lambda} \in \Lambda} q^{\|{\bm \lambda}+{\bm u}_k-{\bm u}_\ell \|^2}.
    \end{IEEEeqnarray}
    For a geometrically uniform packing $\Gamma$, where the set of distances is preserved for every point, then it reduces to
    \begin{IEEEeqnarray}{c}
      \label{eq:theta_equidistance}
      \Theta_\Gamma(z) = \displaystyle\sum_{k=1}^{M}\displaystyle\sum_{{\bm \lambda} \in \Lambda} q^{\|{\bm \lambda}+{\bm u}_k-{\bm u}_1 \|^2}.
    \end{IEEEeqnarray}
  \end{proposition}

  Packings obtained from Construction C are periodic and in particular, a $2$-level Construction C, written as $\code{C}_1 + 2\code{C}_2 + 4\mathbb{Z}^n$, is geometrically uniform, so we can apply Proposition~\ref{prop:thm_thetanonl}, more specifically, \eqref{eq:theta_equidistance}.
  
  In~\eqref{eq:theta_equidistance}, we identify $\boldsymbol{\lambda} \in 4\Integers^n$, $M=|\code{C}_1| |\code{C}_2|$, and $\vect{u}_1=(0,\dots,0)$, since $\code{C}_1, \code{C}_2$ are linear codes and thus contain the zero codeword. Notice that in our context, ${\bm u}_k \in \code{C}_1 + 2\code{C}_2$ and initially, let us fix $k$ and set ${\bm u} ={\bm u}_k$ to simplify.
  
  As ${\bm u} \in \code{C}_1+2\code{C}_2$, there exist ${\bm c}_{1} \in \code{C}_1$ and ${\bm c}_{2} \in \code{C}_2$ such that ${\bm u} = {\bm c}_{1} + 2{\bm c}_{2}$.
  The coordinates of ${\bm u}$ can be $0,1, 2$ or $3$ and their frequency are given respectively by $d_{0,0}({\bm c}_{1}, {\bm c}_{2}),$ $d_{1,0}({\bm c}_{1}, {\bm c}_{2}), d_{0,1}({\bm c}_{1}, {\bm c}_{2})$, and $d_{1,1}({\bm c}_{1}, {\bm c}_{2})$, as in \eqref{eq:def_dij}.
  
  By fixing the $i$-th coordinate of ${\bm u},$ we have as possible exponents of $q$
  in~\eqref{eq:theta_equidistance}  
  \begin{IEEEeqnarray}{c}
    4z_i + {\bm u}_{{i}} = \begin{cases}
    4z_i,                   & \text{if } {\bm u}_{{i}}=0 \\
    4(z_i + \tfrac{1}{4}),  & \text{if} ~ {\bm u}_{{i}}=1 \\
    4(z_i + \tfrac{1}{2}),  & \text{if} ~ {\bm u}_{{i}}=2 \\
    4(z_i + \tfrac{3}{4}),  & \text{if} ~ {\bm u}_{{i}}=3 \\
  \end{cases}.
  \label{eq:couting_4z}
\end{IEEEeqnarray}

The corresponding theta series associated to each one of the previous cases are
\begin{IEEEeqnarray*}{rCl}
  \Theta_{4\mathbb{Z}}(z) = \vartheta_{3}(16z),\, \Theta_{4\big(\mathbb{Z}+ \tfrac{1}{2}\big)}(z) = \vartheta_{2}(16z),
  \nonumber \\
  \Theta_{4\big(\mathbb{Z}+\tfrac{1}{4}\big)}(z) = \Theta_{4\big(\mathbb{Z}+ \tfrac{3}{4}\big)}(z) = \frac{\vartheta_{2}(4z)}{2}, 
\end{IEEEeqnarray*}

By incorporating such results into the fixed $n$-dimensional vector ${\bm u},$ we have that
\begin{IEEEeqnarray*}{c}
  \sum_{{\bm z} \in \mathbb{Z}^n} q^{\|4{\bm z}+{\bm u} \|^2} = \vartheta_3^{d_{0,0}}(16z)  \left(\frac{\vartheta_{2}(4z)}{2}\right)^{{d_{1,0}}+ {d_{1,1}}} \vartheta_2^{d_{0,1}}(16z).
\end{IEEEeqnarray*}

Finally, running through all $k$ vectors ${\bm u}_k$ and considering the scaled version $\PConstrC{\code{C}_1}{\code{C}_2} = \frac{1}{2}(\code{C}_1 + 2\code{C}_2 + 4\mathbb{Z}^n),$ we get
\begin{IEEEeqnarray}{rCl}
  \Theta_{\Gamma_\textnormal{C}}(z) &  = &   \displaystyle\sum_{k=1}^{M}\displaystyle\sum_{{\bm z} \in \mathbb{Z}^n} q^{\|\tfrac{1}{2}(4{\bm z}+{\bm u}_k) \|^2} \nonumber \\
  & = & \sum_{\vect{c}_{1_k} \in \code{C}_1} \sum_{\vect{c}_{2_k} \in \code{C}_2} \vartheta_3^{d_{0,0}}(4z) \left(\frac{\vartheta_{2}(z)}{2}\right)^{{d_{1,0}} + {d_{1,1}}}  \vartheta_2^{{d_{0,1}}}(4z),\nonumber \\
  \label{eq:theta_gammac}\IEEEeqnarraynumspace
\end{IEEEeqnarray}
where 
${\bm u}_k = \vect{c}_{1_k} + 2\vect{c}_{2_k},$ for $\vect{c}_{1_k} \in \code{C}_1$ and $\vect{c}_{2_k} \in \code{C}_2$.
\end{IEEEproof}
}}

Theorem~\ref{thm:theta-series_2-level-constructionC} is general and can be applied to nonlattice packings. 
We relate now the theta series of $\PConstrC{\code{C}_1}{\code{C}_2}$ to jwe.
\begin{corollary}
  \label{coro:theta-series_2-level-ConstructionC}
  The theta series of a $2$-level Construction C lattice, in terms of the jwe of two codes, is
  \begin{IEEEeqnarray*}{c}
    \Theta_{\PConstrC{\code{C}_1}{\code{C}_2}}(z)=\textnormal{jwe}_{\code{C}_1,\code{C}_2}\bigl(\vartheta_3(4z),\vartheta_2(4z), \nicefrac{\vartheta_2(z)}{2},\nicefrac{\vartheta_2(z)}{2} \bigr).\label{eq:Theta-ft_2-level-ConstructionC}\IEEEeqnarraynumspace
  \end{IEEEeqnarray*}
\end{corollary}

If we consider the $\mathbb{Z}_4$-linear code $\code{C}$, the theta series of a Construction $\textnormal{A}_4$ lattice can be expressed as follows.
\begin{corollary}
    \label{coro:theta-series_ConstrAfour}
  Let $\code{C}$ be a $\Integers_4$-linear code with $\textnormal{swe}_{\code{C}}(a,b,c)$, then the theta series of $\ConstrAfour{\code{C}}$ is 
  \begin{IEEEeqnarray*}{c}
    \Theta_{\ConstrAfour{\code{C}}}(z) = \textnormal{swe}_{\code{C}}(\vartheta_3(4z), \nicefrac{\vartheta_2(z)}{2}, \vartheta_2(4z)).
  \end{IEEEeqnarray*}
\end{corollary}

\ifthenelse{\boolean{short_version}}{}{{
\begin{IEEEproof}
If the $\Integers_4$-linear code $\code{C}$ is such that $\code{C}=\code{C}_1+2\code{C}_2$, the result comes immediately from Corollary~\ref{coro:theta-series_2-level-ConstructionC}, since $\textnormal{swe}_{\code{C}}(a,b,c)=\textnormal{jwe}_{\code{C}_1,\code{C}_2}(a,c,b,b)$. For a general $\Integers_4$-linear code $\code{C}$, the same proof of Theorem~\ref{thm:theta-series_2-level-constructionC} can be applied, since $\ConstrAfour{\code{C}} = \code{C}+4\Integers^n$ is also a periodic packing and the coordinates of an element in $\ConstrAfour{\code{C}}$ are also described as in \eqref{eq:couting_4z}. The only difference is that the exponents in \eqref{eq:theta_gammac} are replaced by $n_0({\bm c}), n_1({\bm c})+n_3({\bm c})$, and $n_2({\bm c})$ respectively, and the result follows.
\end{IEEEproof}
}}

We can conclude that the packing obtained from Construction $\textnormal{A}_4$ through $\code{C}_1+2\code{C}_2$ over $\Integers_4$ has exactly the same theta series as the packing constructed by the $2$-level Construction C via $\code{C}_1$ and $\code{C}_2$. Despite this equivalence, both results have their own importance, as Corollary~\ref{coro:theta-series_2-level-ConstructionC} can be applied to any choices of $\code{C}_1,\code{C}_2$ and Corollary~\ref{coro:theta-series_ConstrAfour} is restricted to lattices.

In the end of this section, we highlight the following result.
\begin{corollary}
  \label{cor:FSD_Z4-FUM-lattices}
  If $\code{C}$ is a formally self-dual $\Integers_4$-linear code, then $\ConstrAfour{\code{C}}$ is formally unimodular.
\end{corollary}
This is a direct consequence of Corollary~\ref{coro:theta-series_ConstrAfour}. 

\section{Secrecy Gain of Formally Unimodular Lattices}
\label{sec:secrecy-gain_FU-lattices}


\subsection{The Secrecy Function of a Lattice}
\label{sec:secrecy-function_lattice}
    
We start by the definition of \emph{secrecy gain}~\cite{OggierSoleBelfiore16_1}.
\begin{definition}[Secrecy function and secrecy gain~{\cite[Defs.~1 and~2]{OggierSoleBelfiore16_1}}]
  \label{def:secrecy_function}
  Let $\Lambda$ be a lattice with volume $\vol{\Lambda}=\nu^n$. The secrecy function of $\Lambda$ is defined by
  \begin{IEEEeqnarray*}{c}
    \Xi_{\Lambda}(\tau)\eqdef\frac{\Theta_{\nu\Integers^n}(i\tau)}{\Theta_{\Lambda}(i\tau)},
    \label{eq:def_secrecy-function}
  \end{IEEEeqnarray*} 
  for $\tau\eqdef -i z>0$. The \emph{(strong) secrecy gain} of a lattice is given by $\xi_{\Lambda}\eqdef\sup_{\tau>0}\Xi_{\Lambda}(\tau)$.
\end{definition}

It was shown in~\cite{OggierSoleBelfiore16_1} that the higher the secrecy gain of a lattice, the more security of the lattice wiretap code is. Hence, the objective here is to design a good lattice $\Lambda$ to achieves a high secrecy gain.

Under the design criterion of secrecy function, we summarize the following three important observations for the formally unimodular lattices~\cite{BollaufLinYtrehus22_1app}.
\begin{enumerate}
\item The secrecy function of a formally unimodular lattice $\Lambda$ has exactly the same \emph{symmetry point} at $\tau=1$ as a unimodular or an isodual lattice, i.e., $\Xi_\Lambda(\tau) = \Xi_\Lambda\bigl(\frac{1}{\tau}\bigr)$.
\item Similar to Belfiore and Sol{\'{e}}'s conjecture from~\cite{BelfioreSole10_1}, it is also conjectured that the secrecy function of a formally unimodular lattice $\Lambda$ achieves its maximum at $\tau=1$, i.e., $\xi_{\Lambda}=\Xi_{\Lambda}(1)$.
\item It was demonstrated that formally unimodular lattices can outperform the secrecy gain of unimodular lattices. 
In particular, the unimodular and formally unimodular lattices constructed via Construction A are compared, and it indicates that formally unimodular lattices obtained from formally self-dual codes via Construction A always achieve better secrecy gains than the Construction A unimodular lattices obtained from self-dual codes (see \cite[Tab.~I]{BollaufLinYtrehus22_1app} for details).
\end{enumerate}

Using these observations, we next 
explore the secrecy gain of formally unimodular lattices obtained by Construction $\textnormal{A}_4$ from formally self-dual codes over $\Integers_4$.

\subsection{Secrecy Gain of Construction $\textnormal{A}_4$ Lattices obtained from Formally Self-Dual Codes over $\Integers_4$} 
\label{sec:secrecy-gain_ConstrAfour-lattices_fsd-codes_Z4}
    
In this subsection, we derive a closed-form expression of the theta series of a Construction $\textnormal{A}_4$ lattice obtained from a formally self-dual $\Integers_4$-linear code. Let $\code{C}$ be a $\Integers_4$-linear code. From \eqref{eq:swe-MacWilliams-identity_FSD-codes_Z4} and the following identities from \cite[Eq.~(23), Ch.~4]{ConwaySloane99_1}, and \cite[Eq.~(31), Ch.~4]{ConwaySloane99_1}, respectively,
\begin{IEEEeqnarray}{rCl}
  \vartheta_3(z)+\vartheta_4(z)& = &2\vartheta_3(4z), \vartheta_3(z)-\vartheta_4(z)=2\vartheta_2(4z),\quad\label{eq:useful-identities-1}
  \\[1mm]
  \vartheta^4_2(z)+\vartheta^4_4(z)& = &\vartheta^4_3(z),\quad\label{eq:useful-identities-2}
\end{IEEEeqnarray}
we obtain
\begin{IEEEeqnarray}{rCl}
  \Theta_{\ConstrAfour{\code{C}}}(z)& = &\textnormal{swe}_{\code{C}}(\vartheta_3(4z), \nicefrac{\vartheta_2(z)}{2}, \vartheta_2(4z))
  \nonumber\\
  & \stackrel{\eqref{eq:swe-MacWilliams-identity_FSD-codes_Z4}}{=} &\inv{\bigcard{\dual{\code{C}}}}\textnormal{swe}_{\code{C}}\Big(\vartheta_3(4z)+2\frac{\vartheta_2(z)}{2}+\vartheta_2(4z), \nonumber \\
  & & \vartheta_3(4z)-\vartheta_2(4z), \vartheta_3(4z)-2\frac{\vartheta_2(z)}{2}+\vartheta_2(4z) \Big) \nonumber\\
  & \stackrel{\eqref{eq:useful-identities-1}}{=} &\inv{\bigcard{\dual{\code{C}}}}\textnormal{swe}_{\code{C}}\Big(\vartheta_3(z)+\vartheta_2(z), \vartheta_4(z), \nonumber\\
  & & \vartheta_3(z)-\vartheta_2(z)\Big)
  \nonumber\\
  & \stackrel{\eqref{eq:swe-MacWilliams-identity_FSD-codes_Z4}}{=} &\bigcard{\dual{\code{C}}}^{-2}\textnormal{swe}_{\code{C}}\Big(2\vartheta_3(z)+2\vartheta_4(z), 2\vartheta_2(z),   \nonumber\\
  & & 2\vartheta_3(z)-2\vartheta_4(z)\Big)
  \nonumber\\
  & \stackrel{\eqref{eq:useful-identities-2}}{=} &\bigcard{\dual{\code{C}}}^{-2}\cdot 2^n\textnormal{swe}_{\code{C}}\Bigl(\vartheta_3(z)+\vartheta_4(z), \nonumber\\
  & & \sqrt[4]{\vartheta^4_3(z)-\vartheta^4_4(z)}, \vartheta_3(z)-\vartheta_4(z)\Bigr).
  \nonumber\\
  & = &2^{-n}\cdot\textnormal{swe}_{\code{C}}\Bigl(\vartheta_3(z)+\vartheta_4(z),  \nonumber\\
  & & \sqrt[4]{\vartheta^4_3(z)-\vartheta^4_4(z)}, \vartheta_3(z)-\vartheta_4(z)\Bigr),\label{eq:theta-series_FSD-codes_Z4}
\end{IEEEeqnarray}
where \eqref{eq:theta-series_FSD-codes_Z4} holds since if $\code{C}$ is formally self-dual, $\card{\dual{\code{C}}}=4^{\nicefrac{n}{2}}$.

Now we are able to state the following main theorem.
\begin{theorem}
  \label{thm:inv_secrecy-function_SymmetrizedWeightEnumerator}
  Let $\code{C}$ be a formally self-dual code over $\Integers_4$. Then
  \begin{IEEEeqnarray*}{c}
    \inv{\Bigl[\Xi_{\ConstrAfour{\code{C}}}(\tau)\Bigr]}=\frac{\textnormal{swe}_{\code{C}}\bigl(1+t, \sqrt[4]{1-t^4}, 1-t\bigr)}{2^{n}},\label{eq:Xi-ft_ConstructionA_FSDcodes}\IEEEeqnarraynumspace
  \end{IEEEeqnarray*}
  where $0<t(\tau)\eqdef\nicefrac{\vartheta_4(i\tau)}{\vartheta_3(i\tau)} < 1$. Moreover, define $h_{\code{C}}(t)\eqdef\textnormal{swe}_{\code{C}}\bigl(1+t, \sqrt[4]{1-t^4}, 1-t\bigr)$ for $0< t < 1$. Then, maximizing the secrecy function $\Xi_{\ConstrAfour{\code{C}}}(\tau)$ is equivalent to determining the minimum of $h_{\code{C}}(t)$ on $t\in(0,1)$.
\end{theorem}

\begin{example}
  \label{ex:codes_dim8}
 Consider two formally self-dual codes over $\Integers_4$ in dimension $8$. The first one is the octacode $\code{O}_8$, and its $\textnormal{swe}_{\code{O}_8}$ presented in~\cite[Ex.~14.3]{Wan97_1}.  
 Also, consider the formally self-dual code $\code{C}_8$ from~\cite[pp.~83--84]{BetsumiyaHarada03_1}, one can obtain
 \begin{IEEEeqnarray*}{rCl}
   \IEEEeqnarraymulticol{3}{l}{%
     \textnormal{swe}_{\code{C}_{8}}(a,b,c)}\nonumber\\*\quad%
   & = & c^8 + 64b^8 + 12a b^2 c^5 + 64a b^6 c +16 a^2 c^6\nonumber\\
   && +\> 40a^3 b^2 c^3 + 30a^4 c^4 + 12a^5 b^2 c + 16a^6 c^2 + a^8.
 \end{IEEEeqnarray*} 
 We have that $h_{\code{O}_8}(t) = 256 (1 - t^4 + t^8)$ and $h'_{\code{O}_8}(t) = 256(-4t^3+8t^7)$. When we solve $h'_{\code{O}_8}(t) = 0$, we get as unique solution in the interval $t \in (0,1)$, $t=\nicefrac{1}{\sqrt[4]{2}}$. For $t \in (0, \nicefrac{1}{\sqrt[4]{2}})$, $h'_{\code{O}_8}(t) = 4t^3(-1+2t^4)<0$ and for  $t \in (\nicefrac{1}{\sqrt[4]{2}},1),$ $h'_{\code{O}_8}(t)>0$, meaning that $t=\nicefrac{1}{\sqrt[4]{2}}$ is a minimum, as we wanted. Therefore, $\xi_{\ConstrA{\code{O}_8}} \approx 1.333$, which coincides with the best known secrecy gain up to now in this dimension. 
 
 Proceeding analogously for $\code{C}_8$, $h_{\code{C}_8}(t) = 64 \bigl(2 t^8+t^6-\left(\sqrt{1-t^4}+2\right) t^4 -\left(\sqrt{1-t^4}-1\right) t^2+2 \left(\sqrt{1-t^4}+1\right)\bigr)$ and $h'_{\code{C}_8}(t)=0$ for $t=\nicefrac{1}{\sqrt[4]{2}}$, which is also a minimum. For this code, $\xi_{\ConstrA{\code{C}_8}} \approx 1.282 < \xi_{\ConstrA{\code{O}_8}}$.\hfill\exampleend
\end{example}

\ifthenelse{\boolean{short_version}}{}{{
  \begin{example}
    \label{ex:optimal_codes}
    Gulliver and Harada presented in~\cite{GulliverHarada01_1} optimal formally self-dual codes over $\Integers_4$ in dimensions $6,8,10$ and $14$, together with their swes. Each $h_{\code{C}_i}(t)$, $i=6,8,10,14$ achieves its minimum at $t=\nicefrac{1}{\sqrt[4]{2}}$. Therefore, we have $\xi_{\ConstrAfour{\code{C}_{6}}} \approx 1.172$, $\xi_{\ConstrAfour{\code{C}_{8}}} \approx 1.333$, $\xi_{\ConstrAfour{\code{C}_{10}}} \approx 1.379$, and $\xi_{\ConstrAfour{\code{C}_{14}}} \approx 1.871$, which coincide or are very close to best secrecy gains from~\cite[Tab.~I]{BollaufLinYtrehus22_1app}.\hfill\exampleend
  \end{example}
}}

\begin{example}
  \label{ex:n22_FSD-code_Z4}
  In this example we consider the swe of the isodual code $\code{D}_{4,22}$ presented in~\cite[p.~230, Prop.~4.2]{BachocGulliverHarada00_1}. Using Theorem~\ref{thm:inv_secrecy-function_SymmetrizedWeightEnumerator}, we get $\xi_{\ConstrAfour{\code{D}_{4,22}}}\approx 3.403$. We remark that the best known secrecy gain for formally unimodular lattice in this dimension is $3.34$, which is presented in~\cite[Tab.~I]{BollaufLinYtrehus22_1app}.\hfill\exampleend
\end{example}

\subsection{Secrecy Gain of Construction $\textnormal{A}_4$ Lattices obtained from $\code{C}_1+2\code{C}_2$ $\Integers_4$-Linear Codes}
\label{sec:secrecy-gain_C1plus2C2-Z4-linear-codes}

We review an important class of binary Reed-Muller codes in coding theory.
\begin{definition}[{Reed-Muller codes~\cite[Ch.~13]{MacWilliamsSloane77_1}}]
  \label{def:reed-muller-codes}
  For a given $m\in\Naturals$, the $r$-th order binary Reed-Muller code $\code{R}(r,m)$ is a 
  linear $[n=2^m,k=\sum_{i=0}^r\binom{m}{i}]$ code 
  for $r\in [0:m]$, constructed as the vector space spanned by the set of all $m$-variable Boolean monomials of degree at most $r$.
\end{definition}


Reed-Muller codes have interesting properties, such as being nested. In order to get $\Integers_4$-linear codes from pairs of Reed-Muller binary codes, we still need to guarantee that the chain is closed under the element-wise product, which is true for the chains described next.

A result connecting the construction of $\mathbb{Z}_4$-linear codes and Reed-Muller chains is the following.
\begin{proposition}[{\cite[Ex.~12.8]{Wan97_1}}]
  \label{prop:reedmuller_unimodular}
  The $\mathbb{Z}_4$-linear code $\widebar{\code{C}}_{2^m} \eqdef \code{R}(1,m)+2\code{R}(m-2,m)$ induces a unimodular lattice $\ConstrAfour{\widebar{\code{C}}_{2^m}} = \tfrac{1}{2} \left( \widebar{\code{C}}_{2^m} + 4\mathbb{Z}^{2^m} \right)$. 
\end{proposition}


\begin{example}
  \label{ex:BWs_dims16-32}
  Proposition~\ref{prop:reedmuller_unimodular} gives an even unimodular lattice in dimension $16$ obtained via $\Lambda_{\textnormal{A}_4}(\widebar{\code{C}}_{16})$, where $\widebar{\code{C}}_{16} =  \code{R}(1,4)+2\code{R}(2,4)$ is isomorphic to $\lattice{E}_8 \times \lattice{E}_8$. For such lattice, $\xi_{\ConstrAfour{\widebar{\code{C}}_{16}}} \approx 1.778 < 2.141$, see Table~\ref{tab:table_secrecy-gains_FU-lattices_z4}. If one considers
  \begin{IEEEeqnarray}{c}\label{eq:c32}
    \widebar{\code{C}}_{32}  =  \code{R}(1,5)+2\code{R}(3,5), \label{eq:c32}
  \end{IEEEeqnarray}
  then $\textnormal{BW}_{32}=\sqrt{2}\ConstrAfour{\widebar{\code{C}}_{32}}=\tfrac{\sqrt{2}}{2} \left(\widebar{\code{C}}_{32}+4\mathbb{Z}^{32}\right)$ is an unimodular lattice in dimension $32$.
  
  The code $\widebar{\code{C}}_{32}$ as in~\eqref{eq:c32} is self-dual. Hence, Theorem~\ref{thm:inv_secrecy-function_SymmetrizedWeightEnumerator} can be applied and we get $\xi_{\ConstrAfour{\widebar{\code{C}}_{32}}} \approx 7.11$, which is the best known secrecy gain up to now~\cite[p.~5698]{OggierSoleBelfiore16_1}.\hfill\exampleend
\end{example}


    
\begin{table}[t!]
  \centering
  \caption{Comparison of (strong) secrecy gains of Construction $\textnormal{A}_4$ lattices of even dimensions $n$. 
  }
  \label{tab:table_secrecy-gains_FU-lattices_z4}
  \vskip -2.0ex
  \Scale[0.98]{\begin{IEEEeqnarraybox}[
    \IEEEeqnarraystrutmode
    \IEEEeqnarraystrutsizeadd{3.5pt}{3.0pt}]{V/c/V/c/V/c/V/c/V}
    \IEEEeqnarrayrulerow\\
    & [n, M, d_{\textnormal{Lee}}] 
    && \textnormal{Reference}
    && \xi_{{\Lambda_{\textnormal{A}_4}(\code{C})}}
    && \textnormal{Best-known}~\textnormal{\cite{BollaufLinYtrehus22_1app, OggierSoleBelfiore16_1}}
    &\\
    \hline\hline
    & [6,2^6,4]^{\textnormal{fsd}}  && \textnormal{\cite[p.~125]{GulliverHarada01_1}} && \mathbf{1.172} && \mathbf{1.172} & \\
    \IEEEeqnarrayrulerow \\
    & [8,2^8,6]^{\textnormal{sd}}  && \textnormal{\cite[p.~505]{huffmanpless98_1}} &&  \mathbf{1.333}  && \mathbf{1.333} & \\
    \IEEEeqnarrayrulerow \\
    & [10,2^{10},6]^{\textnormal{fsd}} && \textnormal{\cite[p.~127]{GulliverHarada01_1}}  && 1.379 && \mathbf{1.478} &  \\
     \IEEEeqnarrayrulerow \\
     & [12,2^{12},6]^{\textnormal{fsd}} &&  \textnormal{\cite{ConwaySloane93_1}}  && 1.456 && \mathbf{1.657} & \\
    \IEEEeqnarrayrulerow  \\
    & [12,2^{12},4]^{\textnormal{fsd}} && \textnormal{Ex.~\ref{ex:codes_dim12}}  &&  1.6 && \mathbf{1.657} & \\
    \IEEEeqnarrayrulerow \\
    & [14,2^{14},8]^{\textnormal{fsd}} && \textnormal{\cite[p.~125]{GulliverHarada01_1}}  && 1.871 && \mathbf{1.875} & \\
     \IEEEeqnarrayrulerow \\
    & [16,2^{16},8]^{\textnormal{sd}} && \widebar{\code{C}}_{16},\,\textnormal{Ex.~\ref{ex:BWs_dims16-32}}  && 1.778 && \mathbf{2.141} & \\
     \IEEEeqnarrayrulerow \\
     & [22,2^{22},10]^{\textnormal{fsd}} && \textnormal{\cite[p.~230]{BachocGulliverHarada00_1}} && \mathbf{3.403} &&  3.335  & \\
      \IEEEeqnarrayrulerow \\
    & [24,2^{24},12]^{\textnormal{sd}} && \textnormal{\cite[p.~494]{HuffmanPless03_1}}  && \mathbf{4.063} && \mathbf{4.063} &  \\
      \IEEEeqnarrayrulerow
  \end{IEEEeqnarraybox}}
\end{table}

\section{Formally Self-Dual $\mathbb{Z}_4$-Linear Codes from Nested Binary Codes}
\label{sec:formally-self-dual-Z4-codes}

In this section, we present now a novel construction of formally self-dual codes over $\mathbb{Z}_4$ and investigate the corresponding lattice properties via Construction $\textnormal{A}_4$.

First, we are going to investigate how to generate formally self-dual codes over $\Integers_4$, which by itself is an interesting research topic~\cite{GulliverHarada01_1,BetsumiyaHarada03_1}. 
    


The dual of a $\Integers_4$-linear code $\code{C}=\code{C}_1+2\code{C}_2$ is as follows.
\begin{lemma}
  \label{lem:dual_C1-2C2}
  Let $\code{C} = \code{C}_1 + 2\code{C}_2$ be a $\mathbb{Z}_4$-linear code. Then, $\code{C}^\perp = \code{C}_2^\perp + 2\code{C}_1^\perp$.
\end{lemma}

\begin{IEEEproof}
  First, we notice that $\code{C}_2^\perp \subseteq \code{C}_1^\perp$, which comes from the fact that $\code{C} = \code{C}_1 + 2\code{C}_2$ is linear over $\mathbb{Z}_4$, therefore $\code{C}_1 \subseteq \code{C}_2$ is closed under element-wise product. 
  
  We will demonstrate that $\code{C}_2^\perp + 2\code{C}_1^\perp \subseteq \code{C}^\perp$. Consider an element $\vect{c}_2 + 2\vect{c}_1 \in \code{C}_2^\perp + 2\code{C}_1^\perp$ and $\vect{c}_1 + 2\vect{c}_2 \in \code{C}$. Hence, 
  \begin{IEEEeqnarray}{rCl}
    \langle \vect{c}_2 + 2\vect{c}_1, \vect{c}_1 + 2\vect{c}_2  \rangle & = & \langle \vect{c}_2, \vect{c}_1  \rangle + 2\langle \vect{c}_2, \vect{c}_2 \rangle + 2\langle \vect{c}_1, \vect{c}_1 \rangle  \nonumber \\
    & & + 4 \langle \vect{c}_1, \vect{c}_2 \rangle = 0 \bmod 4.
  \end{IEEEeqnarray}
  Therefore, $\vect{c}_2 + 2\vect{c}_1 \in \code{C}^\perp$. Now, based on the arguments presented in \cite[pp.~33--34]{ConwaySloane93_1}, we observe that $|\code{C}| |\code{C}^\perp| = |\code{C}_1||\code{C}_2||\code{C}_1^\perp||\code{C}_2^\perp| = 2^{k_1}2^{k_2}2^{n-k_1}2^{n-k_2} = 2^{2n} = 4^n$, which is the dimension of $\mathbb{Z}_4^n$ and here, $k_1$ is the dimension of $\code{C}_1$ and $k_2$ is the dimension of $\code{C}_2$. The proof is then complete.
\end{IEEEproof}

It derives immediately from Lemma~\ref{lem:dual_C1-2C2} that if $\code{C}_2 = \code{C}_1^\perp$, then $\code{C}=\code{C}^\perp$ and $\code{C}$ is self-dual.

The result below gives a condition to construct formally self-dual $\Integers_4$-linear codes.    
\begin{theorem}
  \label{thm:fsd-Z4codes_C1plus2C2fsd}
  Let $\code{C} = \code{C}_1 + 2\code{C}_2$ be a $\Integers_4$-linear code. If $W_{\code{C}_1}(x,y) = W_{\code{C}_2^\perp}(x,y)$ and $W_{\code{C}_2}(x,y) = W_{\code{C}_1^\perp}(x,y)$, then $\code{C}$ is formally self-dual. 
\end{theorem}
\ifthenelse{\boolean{short_version}}{}{{
    \begin{IEEEproof}
      The proof is presented in Appendix~\ref{sec:proof-theorem3}.
    \end{IEEEproof}}
}

\begin{example}
  \label{ex:codes_dim12}
  Consider $\code{C}_1$ as the $[12,2,8]$ binary code and $\code{C}_2$ as the $[12,10,2]$ binary code, generated respectively by
  \begin{IEEEeqnarray*}{c}
    \mat{G}_1=\left(\begin{smallmatrix}
      1 & 0 & 1 & 1 & 1 & 0 & 0 & 0 & 1 & 1 & 1 & 1 \\
      0 & 1 & 0 & 0 & 0 & 1 & 1 & 1 & 1 & 1 & 1 & 1
    \end{smallmatrix}\right),\,
    \mat{G}_2=\left(\begin{smallmatrix}
      1 & 0 & 1 & 1 & 1 & 0 & 0 & 0 & 1 & 1 & 1 & 1 \\
      0 & 1 & 0 & 0 & 0 & 1 & 1 & 1 & 1 & 1 & 1 & 1 \\
      0 & 0 & 0 & 0 & 0 & 0 & 0 & 0 & 1 & 1 & 1 & 1 \\
      0 & 0 & 0 & 0 & 0 & 1 & 1 & 1 & 1 & 1 & 1 & 0 \\
      0 & 1 & 1 & 1 & 1 & 0 & 0 & 0 & 0 & 1 & 1 & 1 \\
      0 & 0 & 0 & 1 & 1 & 0 & 0 & 0 & 1 & 1 & 1 & 1 \\
      0 & 0 & 0 & 0 & 0 & 0 & 1 & 1 & 1 & 1 & 1 & 1 \\
      0 & 0 & 0 & 0 & 0 & 0 & 0 & 1 & 1 & 0 & 0 & 0 \\
      0 & 0 & 0 & 0 & 0 & 0 & 0 & 0 & 1 & 1 & 0 & 0 \\
      0 & 0 & 1 & 0 & 1 & 0 & 0 & 0 & 1 & 1 & 1 & 1
    \end{smallmatrix}\right).\IEEEeqnarraynumspace
  \end{IEEEeqnarray*}
  
  Observe that the first two rows of $\mat{G}_2$ correspond to the generators of $\code{C}_1$ and the third is the element-wise product between them. Therefore, we have a guarantee that $\code{C}_1 \subseteq \code{C}_2$ and this chain is closed under the element-wise product. Hence, $\code{C}=\code{C}_1+2\code{C}_2$ is a $\Integers_4$-linear code.
  
  These two codes satisfy the conditions of Theorem~\ref{thm:fsd-Z4codes_C1plus2C2fsd}, i.e.,  $W_{\code{C}_1}(x,y) = W_{\code{C}_2^\perp}(x,y)$ and $W_{\code{C}_2}(x,y) = W_{\code{C}_1^\perp}(x,y)$, but $\code{C}_2 \neq \code{C}_1^\perp$. The swe of $\code{C}$ is 
  \begin{IEEEeqnarray*}{rCl}
    \textnormal{swe}_{\code{C}}(a,b,c)& = & a^{12}+1152 a^2 b^8 c^2+768 a^3 b^8 c
    \nonumber\\
    && +\>192 a^4 b^8 +18 a^{10} c^2+64 a^9 c^3
    \nonumber\\
    && +\>111 a^8 b^4+192 a^7 c^5 +252 a^6 c^6
    \nonumber\\
    && +\>192 a^5 c^7 +111 a^4 c^8+64 a^3 c^9
    \nonumber\\
    && +\>18 a^2 c^{10}+768 a b^8 c^3+192 b^8 c^4+c^{12},
  \end{IEEEeqnarray*}
  which satisfy the MacWilliams identity~\eqref{eq:swe-MacWilliams-identity_FSD-codes_Z4}, hence it is formally self-dual in $\Integers_4$. Moreover, $d_{\textnormal{Lee}}(\code{C})=4$, which is not optimal for this length, but coincides with the best Lee distance of self-dual codes.

  Note that the formally self-dual code in Example~\ref{ex:codes_dim12} has secrecy gain $\xi_{\ConstrAfour{\code{C}_{12}}} \approx 1.6$, which coincides with the performance of self-dual codes. However, it is slightly worse than the best up to now, $1.657$. On the other hand, this example points out an interesting fact: optimal $d_{\textnormal{Lee}}$ does not imply higher secrecy gain: For length $12$, a Lee-optimal code has $d_{\textnormal{Lee}} = 6$. One such code has secrecy gain $\xi_{\ConstrAfour{\code{C}_{12}}} \approx 1.456 < 1.6$, achieved by a $d_{\textnormal{Lee}}= 4$ code.\hfill\exampleend
\end{example}

As a result, we summarize the secrecy gains of some Construction $\textnormal{A}_4$ lattices obtained from formally self-dual codes over $\Integers_4$ in Table~\ref{tab:table_secrecy-gains_FU-lattices_z4}.

\section{Conclusion}
\label{sec:conclusion}

In this work, we studied the secrecy gains of the Construction $\textnormal{A}_4$ lattices from formally self-dual $\Integers_4$-linear codes. We showed that these Construction $\textnormal{A}_4$ lattices are formally unimodular and presented a universal approach to determine their secrecy gains. We found that it is possible to obtain a better secrecy gain from Construction $\textnormal{A}_4$ formally unimodular lattices than that from Construction $\textnormal{A}$ formally unimodular lattices. Furthermore, a novel code construction of formally self-dual $\Integers_4$-linear codes is given. 

\ifthenelse{\boolean{short_version}}{}{{
\appendices


\section{Proof of Theorem~\ref{thm:fsd-Z4codes_C1plus2C2fsd}}
\label{sec:proof-theorem3}

We start the proof by using the following useful identities~\cite[Ch.~5, pp.~148]{MacWilliamsSloane77_1}:
\begin{IEEEeqnarray}{rCl}
  \we{\code{C}_1}(x,y)\we{\code{C}_2}(z,t)& = &\jwe{\code{C}_1}{\code{C}_2}(xz,xt,yz,yt),
  \label{eq:wes-jwe_C1-C2}
  \\
  \jwe{\code{C}_1}{\code{C}_2}(a,b,c,d)& = &\jwe{\code{C}_2}{\code{C}_1}(a,c,b,d).
  \label{eq:swap_jwe_C1-C2}
\end{IEEEeqnarray}
Observe that
\begin{IEEEeqnarray*}{rCl}
  \IEEEeqnarraymulticol{3}{l}{%
    \jwe{\code{C}_1}{\code{C}_2}(xz,xt,yz,yt)}\nonumber\\*\quad%
  & = &\we{\code{C}_1}(x,y)\we{\code{C}_2}(z,t)
  \nonumber\\
  & \stackrel{(i)}{=} &\frac{1}{\card{\dual{\code{C}}_1}}\we{\dual{\code{C}}_1}(x+y,x-y)\frac{1}{\card{\dual{\code{C}}_2}}\we{\dual{\code{C}}_2}(z+t,z-t)
  \nonumber\\
  & \stackrel{(ii)}{=} &\frac{1}{\card{\dual{\code{C}}_1}}\we{\code{C}_2}(x+y,x-y)\frac{1}{\card{\dual{\code{C}}_2}}\we{\code{C}_1}(z+t,z-t)
  \nonumber\\
  & \stackrel{\eqref{eq:wes-jwe_C1-C2}}{=} &\frac{1}{\card{\dual{\code{C}}_2}\card{\dual{\code{C}}_1}}\jwe{\code{C}_2}{\code{C}_1}\bigl(xz+xt+yz+yt,
  \nonumber\\
  && \hspace*{3.00cm}\> xz-xt+yz-yt,\nonumber\\
  && \hspace*{3.00cm}\> xz+xt-yz-yt,\nonumber\\
  && \hspace*{3.00cm}\> xz-xt-yz+yt \bigr)\IEEEeqnarraynumspace
  \nonumber\\
  & \stackrel{\eqref{eq:swap_jwe_C1-C2}}{=} &\frac{1}{\card{\dual{\code{C}}_1}\card{\dual{\code{C}}_2}}\jwe{\code{C}_1}{\code{C}_2}\bigl(xz+xt+yz+xt,
  \nonumber\\
  && \hspace*{3.00cm}\> xz+xt-yz-yt,\nonumber\\
  && \hspace*{3.00cm}\> xz-xt+yz-yt,\nonumber\\
  && \hspace*{3.00cm}\> xz-xt-yz+yt \bigr),\nonumber\IEEEeqnarraynumspace
\end{IEEEeqnarray*}
where $(i)$ follows by the MacWilliams identity and $(ii)$ holds because $W_{\code{C}_2}(x,y) = W_{\dual{\code{C}_1}}(x,y)$ and $W_{\code{C}_1}(z,t) = W_{\dual{\code{C}}_2}(z,t)$. Thus, we have
\begin{IEEEeqnarray}{rCl}
  \IEEEeqnarraymulticol{3}{l}{%
    \jwe{\code{C}_1}{\code{C}_2}(a,b,c,d)}\nonumber\\*\,%
  & = &\frac{1}{\card{\dual{\code{C}}_1}\card{\dual{\code{C}}_2}}\jwe{\code{C}_1}{\code{C}_2}(a+b+c+d,a+b-c-d,\nonumber\\
  && \hspace*{3.00cm}\> a-b+c-d,a-b-c+d).\label{eq:identity_jwe-relation}\IEEEeqnarraynumspace
\end{IEEEeqnarray}
Now, from Proposition~\ref{prop:Z4-linear_ConsC}, Lemma~\ref{lem:dual_C1-2C2}, and the fact that $\swe{\code{C}}(a,b,c)=\swe{\code{C}_1+2\code{C}_2}(a,b,c)=\jwe{\code{C}_1}{\code{C}_2}(a,c,b,b)$, we can further get
\begin{IEEEeqnarray}{rCl}
  \IEEEeqnarraymulticol{3}{l}{
    \frac{1}{\card{\dual{\code{C}}_1}\card{\dual{\code{C}}_2}}\swe{\dual{\code{C}}}(a+c+2b,a-c,a+c-2b)}\nonumber\\*\quad%
  & \stackrel{\eqref{eq:swe-MacWilliams-identity_Z4}}{=} &  \swe{\code{C}}(a,b,c)=\swe{\code{C}_1+2\code{C}_2}(a,b,c)
  \label{eq:use_swe-MacWilliams-identity_Z4}\\
  & = &\jwe{\code{C}_1}{\code{C}_2}(a,c,b,b)
  \nonumber\\
  & \stackrel{\eqref{eq:identity_jwe-relation}}{=} &\frac{1}{\card{\dual{\code{C}}_2}\card{\dual{\code{C}}_1}}\jwe{\code{C}_1}{\code{C}_2}(a+c+2b,a+c-2b,\nonumber\\
  && \hspace*{3.00cm}\> a-c,a-c)
  \nonumber\\
  & = &\frac{1}{\card{\dual{\code{C}}_1}\card{\dual{\code{C}}_2}}\swe{\code{C}}(a+c+2b,a-c,a+c-2b).
  \label{eq:relation_swe-jwe}\IEEEeqnarraynumspace
\end{IEEEeqnarray}
Therefore, by comparing~\eqref{eq:use_swe-MacWilliams-identity_Z4} with~\eqref{eq:relation_swe-jwe}, we obtain $\swe{\code{C}}=\swe{\dual{\code{C}}}$. This completes the proof.
}}

\balance 

\bibliographystyle{IEEEtran}
\bibliography{defshort1,biblioHY}

\end{document}